\documentclass[floatfix,aps,twocolumn,pra,superscriptaddress]{revtex4-1}
\usepackage{graphicx}
\usepackage{dcolumn}
\usepackage{bm}
\usepackage[T1]{fontenc}
\usepackage[latin9]{inputenc}
\usepackage{textcomp}
\usepackage{amsmath}
\usepackage{footnote}
\usepackage{mathrsfs}
\usepackage{graphicx}
\usepackage{amssymb}
\usepackage{units}
\usepackage{xcolor}
\usepackage{footmisc}

\newcommand{\eq}[1]{(\ref{#1})}

\begin{document}

\title{Galilean boosts and superfluidity  of resonantly driven polariton fluids in the presence of an incoherent reservoir}

\author{Ivan Amelio}
\affiliation{INO-CNR BEC Center and Dipartimento di Fisica, Universit{\`a} di Trento, 38123 Povo, Italy}
\author{Anna Minguzzi}
\affiliation{Univ. Grenoble Alpes, CNRS, LPMMC, 38000 Grenoble, France}
\author{Maxime Richard}
\affiliation{Univ. Grenoble Alpes, CNRS, Grenoble INP, Institut N\'{e}el, 38000 Grenoble, France}
\author{Iacopo Carusotto}
\affiliation{INO-CNR BEC Center and Dipartimento di Fisica, Universit{\`a} di Trento, 38123 Povo, Italy}


\begin{abstract}

We theoretically investigate how the presence of a reservoir of incoherent excitations affects the superfluidity properties of resonantly driven polariton fluids. 
While in the absence of reservoir 
the two cases of a defect moving in a fluid at rest and of a fluid flowing against a static defect are linked by a formal
Galilean transformation, here the reservoir defines a privileged reference frame attached to the semiconductor structure and causes markedly different features between the two settings. The consequences on the critical velocity 
for superfluidity are highlighted and compared to
experiments in resonantly driven excitons polaritons.
\end{abstract}

\date{\today}

\maketitle

\section{Introduction}

In the last decade, fluids of exciton-polaritons have emerged as a most powerful platform where to investigate quantum hydrodynamics questions related to superfluidity~\cite{carusotto2013}. A key strength of these systems is that the speed of flow of the fluid can be directly controlled by varying the incidence angle of the pump driving laser, while defects can be engineered both optically and at the sample fabrication stage. 
These features have allowed a direct experimental implementation of the Landau criterion for superfluidity in terms of the density pattern generated by polariton flow past a static defect ~\cite{carusotto2004}.
 In particular, experimental observation of superfluid behaviours was reported~\cite{amo2009}, as well as the hydrodynamic nucleation of vortices~\cite{nardin2011} and solitons~\cite{amo2011}.

In addition to their intrinsic non-equilibrium character~\cite{carusotto2013}, a series of recent experiments has unveiled another novel feature of polariton fluids, namely the presence of a reservoir of incoherent excitations interacting with the polariton fluid and modifying its dynamical properties. While such a reservoir is naturally present under generic incoherent pumping schemes~\cite{wouters2007,bobrovska2014,bobrovska2017dynamical,baboux2018unstable}, its presence was not expected {\em a priori} in coherent pump schemes and has been experimentally established in a series of recent works~\cite{sarkar2010,walker2017,stepanov2019}.

More specifically, our contribution \cite{stepanov2019} inferred the presence of the reservoir from an important modification of the dispersion of collective excitations in the fluid, in particular a significantly reduced sound speed. In order to get a deeper understanding of the role of the reservoir and, at the same time, reconcile our observations with previous works on polariton superflows past a static defect, a general theoretical study of the effect
of Galilean transformations on polariton superfluidity is needed, so to identify the consequences of the privileged frame of reference set by 
the underlying semiconductor cavity structure. These conceptual issues are the subject of the present article.

The structure of the paper is the following. In Sec.~\ref{sec:GalileoBoost} we review the standard theory of Galilean boosts in quantum mechanics, which formally represent a symmetry for the undriven  polariton field within the parabolic approximation for the polariton dispersion;  as a consequence,  these transformations  preserve the  form of the generalized Gross-Pitaevskii equation (GPE) for driven-dissipative polariton fluids 
 in the absence of a reservoir, provided one keeps  into account the proper covariance law for the coherent pump term.
The Galilean transformation is meant as a useful mathematical mapping to link the GPE dynamics of fluids driven with different wavevectors and entails that, 
for a given value of the polariton density and thus of the speed of sound, 
the perturbation induced by a defect only depends on its relative velocity with respect to the fluid.
For instance, our mapping provides a useful link between the relevant situations of polaritons injected against a static defect and of a defect moving in a fluid at rest.
 However, since optics in materials is generally not Lorentz invariant and 
Fresnel drag effects \cite{jackson1975classical,landau1984electrodynamics} take place, such a mapping does not describe
a physical change of frame of reference.
As next step
in Sec.~\ref{sec:LandauCriteria} we provide a reformulation of the Landau criterion for superfluidity in non-equilibrium fluids based on complex-valued wavevectors. In Sec.~\ref{sec:Bogoliubov} we move to the core of our work and we discuss the effect of the reservoir of incoherent excitations on the Bogoliubov dispersion of the collective excitations on top of a spatially homogeneous, coherently pumped polariton fluid. The fact that the incoherent excitations forming the reservoir, e.g. dark excitons, are physically bound to the semiconductor cavity structure,  breaks the formal   
Galilean invariance of the undriven polariton field
and has deep implications for the superfluidity properties of the polariton fluid, as highlighted in Sec.~\ref{sec:LandauReservoir}. The critical relative velocity 
turns out to be different depending on whether a moving polariton fluid is hitting a defect at rest, or a moving defect is flowing through a polariton fluid at rest: indeed, in the former case, the critical velocity is related to the total blue-shift of polariton modes, while in the latter case, it receives the sole contribution of the coherent polariton component.
Conclusions are finally drawn in Sec.\ref{sec:Conclu}.

\section{Galilean Transformations}
\label{sec:GalileoBoost}

The goal of this section is to illustrate a mathematical mapping that relates the GPE dynamics of polariton fluids driven with different pump wavevectors, in the absence of a incoherent reservoir.  
As a corollary,  if   the relative velocities and the polariton densities are the same, the two cases of polaritons injected against a static defect and of a defect moving in a fluid at rest turn out to be the same situation depicted in two different systems of coordinate.
Because of its formal analogy, we call this transformation a Galilean boost, even though it does not correspond to a physical change of reference frame.

We start by considering the standard driven--dissipative Gross--Pitaevskii equation for a resonantly pumped polariton fluid \citep{carusotto2004,carusotto2013},
\begin{multline}
i \partial_t \psi_{lab}(\mathbf{x},t) = \\ = \left( \omega_0 - \frac{\hbar}{2m} \nabla^2 + g |\psi_{lab}|^2 + V_{ext}(\mathbf{x})- i \frac{\gamma}{2} \right) \psi_{lab} + \\ + F_{lab}(\mathbf{x}, t)\,.
\label{eq:unboostedGPE}
\end{multline}
In the conservative part of the evolution, $\omega_0$ is the bottom of the lower polariton band and $m$ is the polariton mass in the parabolic band approximation, $V_{ext}(\mathbf{x})$ the static potential acting on the polaritons, $g$ quantifies the strength of 
polariton--polariton interactions. Concerning the driven-dissipative terms, $\gamma$ is the loss rate and the driving term $F_{lab}(\mathbf{x}, t)$ is proportional to the spatio-temporal profile of the coherent laser amplitude. In particular,  the cavity material  determines the mass and  interactions of the polaritons, but no reservoir of incoherent excitations is assumed to be present at this stage.

Now we develop the aforementioned formal Galilean boost and with some abuse of notation employ the terminology commonly adopted for physical Galilean transformations. 
For instance, when within this analogy we refer to the ``frame of reference moving at velocity $\mathbf{v}_G$ with respect to the lab'', we have in mind  the change of coordinates $\mathbf{y} = \mathbf{x} - \mathbf{v}_G t$ (the lab is just one chosen frame), while  the time variable remains the same in the two coordinate systems.  The chosen convention for the sign of $\mathbf{v}_G$ is such that a fluid moving at velocity $\mathbf{v}_G$ in the laboratory frame is seen as at rest in the boosted one.  
The Galilean transformation is given by the unitary operator
\begin{equation}
\hat{U}_{lab \to G} = e^{i \frac{\hat{\mathbf{p}}\mathbf{v}_G t - \hat{\mathbf{x}}m \mathbf{v}_G}{\hbar} } = e^{i \frac{\hat{\mathbf{p}}\mathbf{v}_G t}{\hbar}}
e^{-i \frac{\hat{\mathbf{x}}m \mathbf{v}_G}{\hbar}} e^{+i \frac{m{v}_G^2 t}{2\hbar}},
\label{eq:GalileoOperator}
\end{equation}
so that applied on the wavefunction reads
\begin{multline}
\psi_{G}(\mathbf{y}) = [\hat{U}_{lab \to G} \psi_{lab} ] (\mathbf{y}) = \\
= e^{-i \frac{m{\mathbf{y}} \mathbf{v}_G}{\hbar}} e^{-i \frac{m{v}_G^2 t}{2\hbar}} 
\psi_{lab}(\mathbf{y}+\mathbf{v}_G t)\,.
\end{multline}

Galilean invariance of the conservative part of the GPE evolution \eq{eq:unboostedGPE} is then guaranteed by the parabolic form of the kinetic energy according to elementary quantum mechanics~\cite{cohen1991quantum}, so that in our  terminology  the (undriven) polariton field is Galilean invariant (instead, the covariance of the driving term is to be discussed in a moment). For usual polariton systems resulting from the strong coupling of a cavity photon mode to an excitonic transition this parabolic approximation is accurate for the typical flow speeds  considered in the experiments~\cite{carusotto2013}.

Concerning the pump and loss terms, it is straightforward to see \footnote{A possible derivation involves the Hamiltonian that generates the GPE \eq{eq:unboostedGPE}; in particular the pump term reads $\hat{H}^{drive}_{lab} = \int d\mathbf{x}
 F_{lab}(\mathbf{x},t) \hat{\Psi}^{\dagger}(\mathbf{x}) + {\textrm h.c.}$, which transforms as $\hat{H}^{drive}_{G} = \hat{U}_{lab \to G} \hat{H}^{drive}_{lab} \hat{U}^{\dagger}_{lab \to G}  
  = \int d\mathbf{y}
 F_{lab}(\mathbf{y}+\mathbf{v}_G t,t) 
 e^{-i \frac{m{\mathbf{y}} \mathbf{v}_G}{\hbar}} e^{-i \frac{m\mathbf{v}_G^2 t}{2\hbar}} 
 \hat{\Psi}^{\dagger}( \mathbf{y}) + \textrm{h.c.}~$} that the dynamics in the boosted frame is described by the same GPE
 \begin{multline}
 i  \partial_t \psi_G(\mathbf{y},t) = \\ = \left( \omega_0 - \frac{\hbar}{2m} \nabla^2 + g |\psi_G|^2 +
 V_{ext}(\mathbf{y} + \mathbf{v}_G t)
 - i \frac{\gamma}{2} \right) \psi_G + \\
 +F_{G}(\mathbf{y}, t),
 \label{eq:boostedGPE}
 \end{multline}
provided the pump term is covariantly transformed according to
\begin{equation}
F_{G}(\mathbf{y},t)= F_{lab}(\mathbf{y} + \mathbf{v}_G t, t) e^{-i \frac{m{\mathbf{y}} \mathbf{v}_G}{\hbar}} e^{-i \frac{m\mathbf{v}_G^2 t}{2\hbar}}.
\label{eq:Fboost}
\end{equation}
Note that this transformation involves a  shift of the wavevector proportional to the velocity as well as an overall frequency shift ${m{v}_G^2 t}/{(2\hbar)}$. The loss term remains unchanged thanks to the spatio-temporally local form that we have assumed from the outset.

Eqs.~\eqref{eq:unboostedGPE} and \eqref{eq:boostedGPE} describe the same dynamics, the observables in the two frames being linked by the usual Galilean  prescriptions:  $|\psi_{lab}(\mathbf{x}, t)|^2 = |\psi_{G}(\mathbf{y}, t)|^2$ for the density and $\mathbf{v}^{flow}_{lab}(\mathbf{x}, t) = \mathbf{v}^{flow}_{G}(\mathbf{y}, t) + \mathbf{v}_{G}$ for the flow velocity, defined in terms of the wavefunction as usual as $\mathbf{v}^{flow}=\hbar\, \textrm{Im}[\psi^* \nabla \psi]/(2m\,|\psi|^2)$.

Along these lines, it is natural to define for a plane-wave \footnote{For finite pump spots the Galilean arguments still hold and p.e. they relate a stationary spot centered at wavevector $\mathbf{k}_p$ with a spot moving with velocity $-\hbar \mathbf{k}_p/m$ but at a given time centered at zero wavelength. The second situation is not so relevant for our discussion, so we always consider pump spots that can be considered spatially uniform (apart from a plane wave factor).} coherent drive with frequency $\omega_p$ and wavevector $\mathbf{k}_p$ the detuning in the frame comoving with the fluid at $\mathbf{v}_p= \hbar \mathbf{k}_p/m$ as $\Delta_{p} = \omega_p - \hbar {k}_p^2/2m - \omega_0$. 
When $\Delta_p = gn_0$ or equivalently $\omega_p = \omega_0 + \hbar {k}_p^2/2m +gn_0 $, the fluid of density $n_0=|\psi_0|^2$ is characterized by a sonic dispersion with speed of sound $mc_s^2 = \hbar \Delta_p = \hbar g n_0$. In what follows, most of the plots will refer to this most remarkable sonic case. Note that the overall phase factor in Eq.~\eqref{eq:Fboost} is needed to ensure that the density and thus the linearity of the dispersion is independent of the Galilean frame.

Before proceeding, it is important to stress that the Galilean transformation discussed here is useful to mathematically relate the dynamics of fluids injected with different speed, but it does {\em not} correspond to a physical change of reference frame, for instance an experimentalist running parallel to the cavity mirrors.
Indeed, for a medium with a refractive index different from unity, light propagation in a  {\em physically} boosted  Lorentz frame at velocity $\mathbf{v}_L$ is affected by the celebrated Fresnel drag effect~\cite{jackson1975classical,landau1984electrodynamics, jones1972,artoni2001,carusotto2003}, which changes the dispersion relation to 
the same order  in $v_L/c$ 
as the Doppler shift:

\begin{equation}
\omega'(\mathbf{k}') = \omega(\mathbf{k}') + \left( \frac{1}{n_{cav}^2}-1 \right) \mathbf{v}_L \cdot \mathbf{k}'+...
\end{equation}
where the apex refers to quantities measured in the Lorentz frame and  $n_{cav}$ is the refractive index inside the cavity.

\section{Landau criterion for superfluidity in non-equilibrium systems}
\label{sec:LandauCriteria}

According to the 
Landau criterion~\cite{landau1981statistical,pitaevskii2016},
a superfluid is able to flow without friction at speed $\mathbf{v}$ around a static defect until it is  energetically favourable to create excitations in  it,  ie  if $\omega(\mathbf{k})+\mathbf{k}\cdot \mathbf{v} \ge 0 $, where $\omega(k)$ is the excitation dispersion of the fluid at rest.
This provides the well-known 
expression for the critical velocity,

\begin{equation}
v_{c}=\min_{\mathbf{k}} \frac{\omega(\mathbf{k})}{k} .
\label{eq:vcr}
\end{equation}

For a weakly interacting fluid of bosons with contact interactions and  dispersion $\omega(k)=c_s |k|$ this  gives $v_{c}=c_s=\sqrt{\hbar\,g\,n_0/m}$ where  $n_0=|\psi_0|^2$ is the particle density, $m$ is the mass and  $g$ the interaction coupling strength.

Alternatively,
a 
weak
defect is able to move through a superfluid without friction
if the dispersion $\omega(\mathbf{k})$ of elementary excitations in the latter has no intersection with a straight line $\omega=-\mathbf{v}\cdot \mathbf{k}$.
For particles with a parabolic dispersion and local interactions, equivalence of the two points of view is ensured by the Galilean invariance.

This formulation applies well to
 superfluids of material particles with a  long  lifetime of the collective excitations, i.e.  where
 the imaginary part  $\textrm{Im}[\omega(\mathbf{k})]$
 of the dispersion relation
 is much smaller than the real part  $\textrm{Re}[\omega(\mathbf{k})]$
 and can be neglected at long wavelength.
However, subtleties arise in the case of driven-dissipative fluids, e.g. the
polariton ones where the real and imaginary parts of the dispersion $\omega(\mathbf{k})$ may have comparable magnitudes.
The crucial importance of this effect for incoherently pumped polariton condensates was first unveiled in~\cite{wouters2010}, where generalized forms of the Landau criterion for driven-dissipative systems were introduced in terms of real frequencies and complex momenta.
In particular, for a polariton superfluid flowing against an obstacle  it was shown that a pattern forms when the velocity of the fluid  is larger than the critical velocity, thus showing clear superfluid-like features even in the presence of drive and dissipation.
For the case of coherent pumping, a phenomenological way of assessing superfluidity is  by computing the drag force, for which a pioneering discussion of the effect of drive and losses  was reported in~\cite{berceanu2012}.
Notice that driven dissipative polariton fluids may not satisfy other
definitions of superfluidity
e.g. the one in terms of the transverse current-current response~\cite{juggins2018}.

In spite of the complications arising from its driven-dissipative nature, the Galilean invariance argument stating that  the critical (relative) speed  depends on the density but not on the reference frame  remains true for the polariton field, as a corollary of what shown in the previous section; importantly, this does {\em not} assume to work at the acoustic point.
More precisely, in the approximation of an infinite  uniform excitation spot\cite{Note2}, the pattern created by scattering against a defect (hence the superfluidity threshold) depends only on density and on the  relative velocity between the defect and the fluid:  
the flow against a static defect and the defect moving through the fluid at rest correspond to the same dynamics viewed in two different Galilean  frames.
Having argued the validity of our next results in any reference frame, in this Section 
we turn to a formal study of the density pattern created when a polariton fluid is coherently  excited into motion against a static obstacle  by a monochromatic pump of frequency $\omega_p$ and in-plane wavevector $\mathbf{k}_p$~\cite{Note2}, with a special focus on the effects due to drive and dissipation.

\begin{figure}[tbp]
\includegraphics[width=0.8\columnwidth]{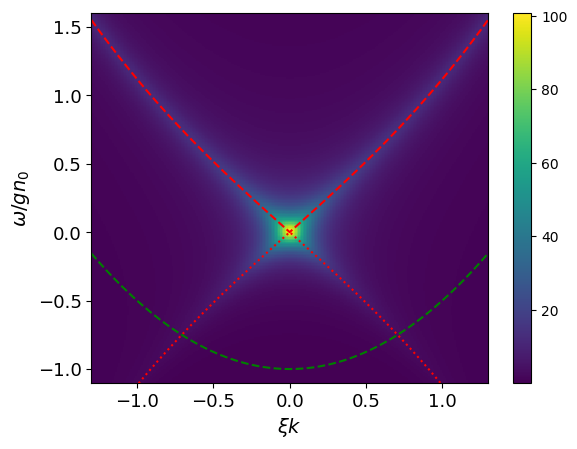}
\caption{Frequency and momentum dependence of the (non-normalized) transmittivity of a coherently pumped cavity to a weak monocromatic probe of frequency $\omega$ (measured with respect to the pump frequency) and wavevector $\mathbf{k}$. The pump injects a coherent polariton fluid at $\mathbf{k}_p=0$ and no incoherent reservoir is assumed to be present. The pump intensity is adjusted to be at the resonant point at which $\Delta_p=gn_0$ and the Bogoliubov dispersion is sonic and gapless. As usual~\cite{wouters2009probing}, the transmission amplitude 
is set by the matrix element $\left.[\omega - \mathscr{L}_{\mathbf{k}}]^{-1}\right|_{11}$ with the Bogoliubov matrix $\mathscr{L}_{\mathbf{k}}$ being defined in Eq.~(\ref{eq:L2x2}). The red dashes (dots) indicate the particle (hole) branch of the Bogoliubov dispersion of the elementary excitations, while the green dashes indicate the bare polariton band at linear regime. Frequencies are measured in units of the interaction energy $g n_0$; lengths (momenta) in units of the (inverse) healing length  $\xi = [{\hbar^2}/{\sqrt{m g n_0}}]^{1/2}$.
}
\label{fig:solita}
\end{figure}

\begin{figure*}[htbp]
\includegraphics[width=2.\columnwidth]{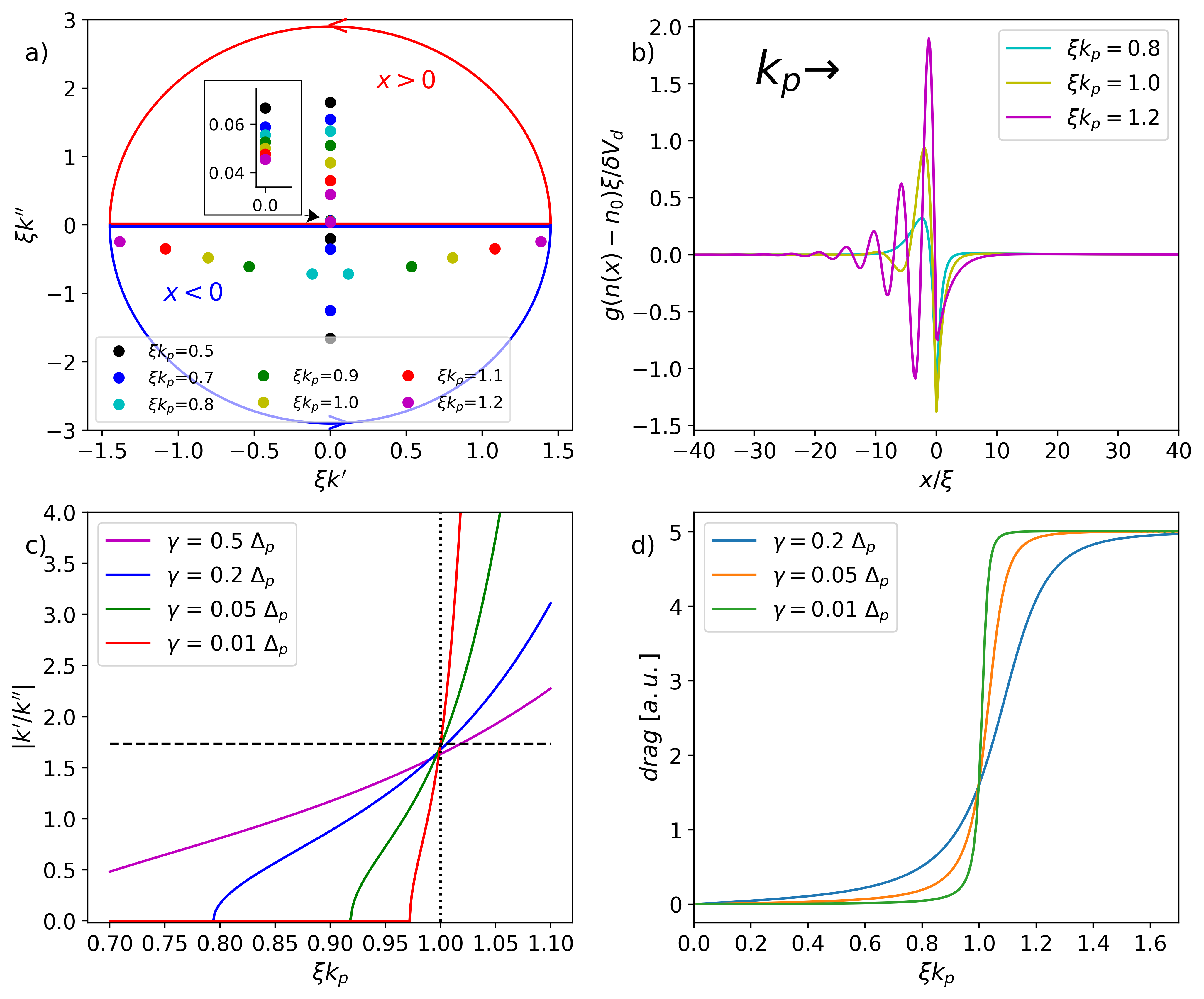}
\caption{(a) Evolution of the four  complex poles of the one-dimensional response function $\chi_{_V}(k\in \mathbb{C})$ for different values of the flow speed $k_p>0$ . The residue theorem is to be applied to the upper (lower) half--plane for $x>0$ ($x<0$).
(b) Spatial profile of the density modulation induced by the defect (located at $x=0$) for different values of $k_p$. More precisely, the renormalized density perturbation $\frac{g[n(x)-n_0]}{\delta V_{def}}$ is reported. (c) Ratio $\textrm{Re}[k]/\textrm{Im}[k]$ between the real and imaginary part of the poles as a function of $k_p$ and for different dissipation rates; the horizontal dashed line indicates $\sqrt{3}$. (d) Drag force as a function of the flow speed for different loss rates $\gamma$.
Across (a--d) the pump frequency is kept at the sonic resonance point $\Delta_p=g n_0$, unless differently specified the damping is set to $\gamma / g n_0=0.2$, and no incoherent reservoir is present. }
\label{fig:1Dsuper}
\end{figure*}

In order to determine the density modulation pattern of the fluid flowing around a weak defect at rest, we 
adopt the method
of Ref.~\onlinecite{carusotto2004} and linearize the GPE \eq{eq:unboostedGPE} on top of the homogeneous solution at the pump's wavevector $\mathbf{k}_p$ and frequency $\omega_p$, via the Ansatz 
\begin{equation}
\psi(\mathbf{x},t) =e^{i (\mathbf{k}_p\cdot\mathbf{x}-\omega_p t)} \left[ \psi_0 + \int d\,\mathbf{ k }\, \  \delta\psi_{\mathbf{ k }} e^{i\mathbf{ k }\cdot\mathbf{x}} \right]\,.
\label{eq:Fourier}
\end{equation}
Importantly, since the defect produces a static
perturbation on the fluid, the wavefunction \eq{eq:Fourier} keeps a monochromatic form at the pump frequency $\omega_p$.

Inserting this Ansatz into the GPE \eq{eq:unboostedGPE} and expanding to 
lowest order in the defect's potential $V_{\rm def}$, one finds for each Fourier component $\delta \psi_{\mathbf{ k }}$
\begin{multline}
\left( \omega_0 + \frac{\hbar (\mathbf{k}_p + \mathbf{ k })^2}{2m} + 2 g |\psi_0|^2 - \omega_p - i \frac{\gamma}{2} \right) \delta \psi_{\mathbf{ k }} + \\ + \psi_0^2  \, \delta \psi_{-\mathbf{ k }}^* = - \delta V_{\rm def}({\mathbf{ k }}) \psi_0\,.
\label{eq:linearized_GPE}
\end{multline}
Then, by combining this equation with the complex conjugate one and solving the matrix inversion problem, we obtain for an arbitrary (real) potential
\begin{equation}
\delta\psi_{\mathbf{ k }} = \chi_{_V}(\mathbf{ k }) \psi_0 \delta V_{\rm def}(\mathbf{ k })
\label{eq:deltapsik}
\end{equation}
where the 
  response function to an external potential is defined as
\begin{equation}
\chi_{_V}(\mathbf{ k }) = [\mathscr{L}(\mathbf{ k })]^{-1}_{11} -  [\mathscr{L}(\mathbf{ k })]^{-1}_{12}
\end{equation}
and  the $2 \times 2$ Bogoliubov matrix reads
\begin{multline}
\mathscr{L} (\mathbf{ k }) = \\
  \begin{pmatrix}
\eta(\mathbf{k_p} + \mathbf{ k }) + gn_0 - i { \gamma}/{2}   &  {g} {\psi_0}^2  \\
- {g} {{\psi_0}^*}^2 & - \eta(\mathbf{k_p} -  \mathbf{ k }) - gn_0 - i { \gamma}/{2}
\end{pmatrix}\,.
\label{eq:L2x2}
\end{multline}
Here, $ \mathbf{k} = \mathbf{k}_p + \mathbf{ k }$ is the (real-valued) total momentum and the detuning function
\begin{equation}
\eta(\mathbf{k}) = \omega_0 +  \hbar k^2/2m +  {g} n_0-\omega_p\,.
\label{eq:eta}
\end{equation}
The perturbed wavefunction in real space is finally obtained from \eq{eq:deltapsik}  by means of the Fourier transform \eq{eq:Fourier}. The Bogoliubov dispersion relation $\omega(\mathbf{k})$ corresponds to the eigenvalues of the matrix \eq{eq:L2x2}, i.e. the zeros of $\det \left[ \omega - \mathscr{L} (\mathbf{ k })  \right]$.

Because of the presence of the imaginary loss terms in \eq{eq:linearized_GPE} and then \eq{eq:L2x2}, the frequency $\omega(\mathbf{k})$ has complex values for real values of the momentum $\mathbf{k}$. Physically, this corresponds to plane-wave excitations having a finite lifetime.
In Fig.~\ref{fig:solita} we plot in red dashes the  dispersion of an interacting polariton fluid at $\mathbf{k}_p=0$ and in the sonic regime. The underlying color plot shows the magnitude of the response $\delta \psi(\mathbf{k}, \omega)$ to a monochromatic optical probe at $(\mathbf{k}, \omega_p + \omega)$, which is proportional to the $(1,1)$ component of the susceptibility matrix, $\left.\left[ {\omega - \mathscr{L}(\mathbf{k})} \right]^{-1}\right|_{11}$. The linewidth of the response along the frequency axis is set by the imaginary part of $\omega$, which in this case is flat and equal to $\gamma$.

While these complex frequency modes allow to study dynamical excitations, it was first noted in~\cite{wouters2010} that the response  of the steady state to static external perturbations is most conveniently characterized
in terms of modes with a real frequency and a complex momentum. In classical electromagnetism~\cite{Tait}, such waves naturally appear when dealing with monochromatic light incident on an absorbing medium. In our context, this point of view is implicitly assumed upon using the residue theorem to evaluate the Fourier integral \eq{eq:Fourier}: since the defect is at rest, it generates a static perturbation in the fluid at $\omega=0$, whose peak wavevectors $\mathbf{ k }$ are determined by the zeros of $\det \mathscr{L} (\mathbf{ k })$, i.e.  the poles of $\chi_{_V} (\mathbf{ k })$.

In the simplest case of a one-dimensional geometry and a delta-like potential at rest giving a momentum-independent $\delta V_{\rm def}(k)=\delta V_{\rm def}$, the position of the poles in the complex $k$ plane are shown in Fig.~\ref{fig:1Dsuper}(a) for different values of the pump wavevector $k_p>0$ (that is, of the speed $\hbar k_p/m$ of the fluid) and a resonant laser frequency $\omega_p=\hbar k_p^2/(2m) + g n_0 +\omega_0$ such that the Bogoliubov dispersion is gapless and has a sonic behaviour with a well-defined speed of sound $c_s$.

With the residue theorem technique, evaluation of the Fourier integral \eq{eq:Fourier} for $x>0$ ($x<0$)  only picks the poles in the upper (lower) complex half-plane. For the $x>0$ region, a single pole is present and this has a vanishing real part. It corresponds then to the exponentially decaying perturbation that is visible in Fig.\ref{fig:1Dsuper}(b) in the $x>0$ downstream region. The faster the flow, the closer the pole to the real axis, so the slower the exponential decay.

The behaviour is richer in the $x<0$ upstream region: for small speeds $k_p$, the two poles have again a vanishing real part and the perturbation displays a monotonic decay. Around $\hbar k_p/mc_s\simeq 0.75$  the poles merge
in the complex $k$ plane at a finite $\textrm{Im}[k]$ and then separate again along a direction parallel to the real axis. For sufficiently large speeds, their real part exceeds the imaginary one so that the perturbation in the fluid starts displaying a clear oscillatory character upstream of the defect.

As the association between the real part of the wavevector and the transferred momentum suggests, this change in behaviour is expected to result into a sharp change in the value of the drag force exerted by the moving fluid onto the defect, defined as~\cite{astrakharchik2004,wouters2010,berceanu2012}
\begin{equation}
 F_d=-\int\,dx \,\nabla V_{\rm def}(x)\,|\psi(x)|^2\,.
 \label{eq:Fd}
\end {equation}

A plot of $F_d$ as a function of the fluid speed is shown in Fig.\ref{fig:1Dsuper}(d) for the sonic case and qualitatively agrees with this prediction.

In particular, the position of the threshold position is consistent with the naive Landau criterion based on comparing the flow speed with the speed of sound. The velocity-independent value of the friction force at high speeds is typical of one-dimensional superfluids and was first anticipated in~\cite{astrakharchik2004} for conservative atomic systems. Finally, the smaller the loss rate $\gamma$, the sharper the transition from a frictionless superfluid behaviour at slow speeds to a finite friction force at fast speeds.

While this picture is qualitatively accurate, establishing a precise relation between the location of the threshold and the behaviour of the poles in the $k$-complex plane requires a bit more careful analysis. As one can see in the lower half-plane of panel (a), the $k$ vectors aquire a real part in fact at a  smaller value $k_p\approx 0.75$ than the threshold that is visible in the force plot around $k_p\approx 1$. To explain this feature, one can see in panel (c) that the different curves of $\textrm{Re}[k]/\textrm{Im}[k]$ for different values of the loss rate cross at a single value close to $\sqrt{3}$ 
\footnote{
Indeed, at $\hbar k_p\simeq m c_s\simeq 1$ the pole condition in terms of  $z=k/k_p$  simply reads $z^4 +iaz + a^2$, whose solution satisfy $|\textrm{Re}[z]/\textrm{Im}[z]| \to \sqrt{3}$ as $a \equiv \frac{\hbar \gamma}{m c_s^2} \to 0$.
}
for a value of the pump wavevector $\hbar k_p\simeq m c_s\simeq 1$ that approximately corresponds to the threshold for the drag force shown in panel (d). This suggests that the threshold is not  determined by the point where the $k$ vectors aquire a real part, but rather by the point when the real part exceeds (by a factor $\sqrt{3}$) the imaginary part.

\section{Bogoliubov dispersion in the presence of an incoherent reservoir}
\label{sec:Bogoliubov}

\begin{figure*}[htbp]
\includegraphics[width=1.8\columnwidth]{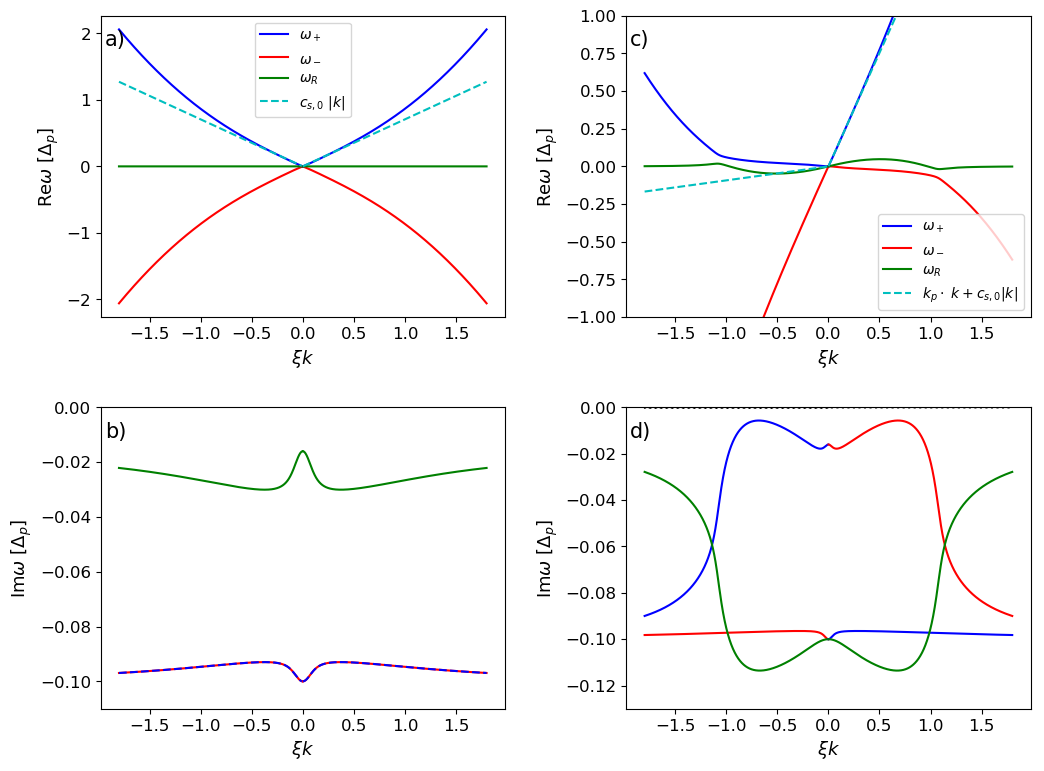}
\caption{Dispersion of collective excitations in a polariton fluid at rest $k_p=0$ [left column, panels (a,b)] and in motion at 
$v_p=0.8 c_{s,T}$ along the $x$ direction [right column, panels (c,d)] in the presence of an incoherent reservoir. Upper (a,c) panels show the real part of the dispersion, the lower (b,d) panels show the imaginary part. The total blue-shift $\mu_T$ is the same in all panels and pumping is tuned at the resonance point such that $\Delta_p=\mu_T$. Other parameters: $\gamma/\mu_T=0.2$, $g_R=2g$ and $\gamma_R = 2\gamma_{inc}=0.08\,\gamma$
which means $c_{s,0} = c_{s,T}/\sqrt{2}$ and $c_s \simeq 0.9 c_{s,0}$. Note that for a slightly larger $\gamma_{inc} \simeq 0.05 \gamma$  or for $v_p \simeq 0.9 c_{s,T}$, the flow configuration in the right panels would become dynamically unstable. The different curves are colored according to their nature at large wavevector $k$. The dashed cyan lines in the upper panels indicate the low-$k$ sonic dispersions \eq{eq:sonic} and \eq{eq:dopplersonic}.
}
\label{fig:disp_kp}
\end{figure*}

The discussion in the previous sections was based on a driven-dissipative, yet purely coherent dynamics of the polariton fluid. Recent experimental works~\cite{sarkar2010,walker2017,stepanov2019} have suggested that an incoherent reservoir of excitations -- most likely of dark-excitonic nature --
is excited
even  under a coherent pump via non-radiative absorption processes.

As it was introduced in these works, the effect of the incoherent reservoir can be theoretically described by including the reservoir density $n_R(\mathbf{x})$ to the equations of motion,
\begin{eqnarray}
i \partial_t \psi &=& \left( \omega_0 - \frac{\hbar \nabla^2}{2m}  + g |\psi|^2 + + g_R n_R - i \frac{\gamma}{2} \right) \psi +  \nonumber\\ &+& F(\mathbf{x},t), \label{eq:psi}\\
\partial_t n_R&=& - \gamma_R n_R + \gamma_{inc} |\psi|^2\,.
\label{eq:n_R}
\end{eqnarray}
Here, the decay of coherent polaritons into incoherent excitations occurs at a rate $\gamma_{inc}$ and the latter give a contribution $g_R n_R$ to the polariton blue-shift. The total decay of polaritons $\gamma$  thus includes a $\gamma_{inc}$ contribution, while the incoherent excitations decay at a rate $\gamma_R$.

At stationarity under a monochromatic pump at $\omega_p$, one has $\psi(\mathbf{x},t)=\psi_0\,\exp(-i\omega_p t)$ and, from \eq{eq:n_R}, one gets a time-independent $n_{R}(\mathbf{x}) =  \frac{\gamma_{inc}}{\gamma_R} |\psi_0(\mathbf{x})|^2$. Reinjecting this expression into \eq{eq:psi}, one simply obtains a renormalized nonlinear coupling strength
\begin{equation}
g_{\rm eff} =g + \frac{ \gamma_{inc}}{\gamma_R} g_R\,.
\label{eq:geff}
\end{equation}
Except for the reinforced interactions and the consequently reinforced blue shift $\mu_T=g_{\rm eff} n_0=gn_0+g_Rn_R$, the reservoir has thus no effect on the stationary state. The usual optical bistability and optical limiting behaviours are found depending on whether the laser frequency $\omega_p$ is blue- or red-detuned as compared to the polariton mode at $\mathbf{k}_p$.

Even more importantly for our purposes, superfluidity features the usual behaviours with a speed of sound defined by the total blue-shift as $mc_{s,T}^2=\mu_T=gn_0+g_Rn_R$. Since this reasoning requires stationarity of both the polariton $\psi(\mathbf{x})$ and the reservoir $n_R(\mathbf{x})$ densities, this result only holds for static defects that do not induce time-dependent modulations to the fluid density, that is defects at rest in a (possibly moving) fluid. And, of course, these statements are only relevant if the fluid is indeed able to reach a dynamically stable steady state: as it was pointed out in \cite{amelio2019a}, the presence of a slow reservoir can in fact give rise to dynamical instabilities that destabilize the stationary state.

The physics gets even more intriguing as soon as one looks at the dynamics of the excitations on top of the fluid, as first noticed in Ref.\onlinecite{stepanov2019}. In the homogeneous case under a plane-wave coherent pump of wavevector $\mathbf{k}_p$ and frequency $\omega_p$, the steady-state solution has the form $\psi_0(\mathbf{r},t) = \psi_0 \exp[i (\mathbf{k_p} \cdot \mathbf{r}-\omega_p t)]$ and the Bogoliubov theory involves a 3$\times$3 matrix
\begin{multline}
\mathscr{L} (\mathbf{ k }) = \\
  \begin{pmatrix}
\eta(\mathbf{k}_p+\mathbf{ k }) + gn_0 - i \frac{ \gamma}{2}  &  {g} {\psi_0}^2 & g_R \psi_0 \\
- {g} {{\psi_0}^*}^2 & - \eta(\mathbf{k}_p-\mathbf{ k }) - gn_0 - i \frac{ \gamma}{2} & -g_R \psi_0 \\
i \gamma_{inc} \psi_0& i \gamma_{inc} {\psi_0}^* & - i \gamma_R
\end{pmatrix},
\label{L3x3}
\end{multline}
where $\mathbf{ k }$ is again the relative wavevector of the excitation on top of the moving fluid and the effective detuning function is now $\eta(\mathbf{k}) = \omega_0 + \hbar k^2/2m +  {g} n_0  + g_R n_R-\omega_p$. The first and second columns/rows of $\mathscr{L} (\mathbf{ k })$ correspond to the polariton modulation $\delta\psi_{\mathbf{ k }}$ and $\delta\psi^*_{-\mathbf{ k }}$, while the third column/row corresponds to the modulation of the reservoir density $\delta n_R$.

The corresponding eigenvalue problem  can be formulated in a physically trasparent way by defining a frequency--dependent effective coupling
\begin{equation}
g_{\rm eff}(\omega) =g + \frac{ \gamma_{inc}}{-i\omega + \gamma_R} g_R ,
\end{equation}
which allows to eliminate the reservoir by replacing  $g$ with $g_{\rm eff}(\omega)$ ~\footnote{Of course, the replacement must not be applied to the $g$ appearing in the definition of the detuning function  $\eta$, which involves the stationary background.}, and thus reduce matrix (\ref{L3x3}) to an effective $2\times 2$ matrix involving the polaritons only. The eigenvalue equation for the collective mode dispersion then reads
\begin{equation}
\left( \omega - \frac{\hbar \mathbf{k}_p}{m}\cdot \mathbf{ k }  + i\frac{\gamma}{2} \right)^2 =
\eta(\tilde{k}) \left[ \eta(\tilde{k}) + 2 g_{\rm eff}(\omega) n_0    \right]
\end{equation}
with $\tilde{k} = \sqrt{k_p^2 + k^2}$. While this expression is formally nearly identical to the usual one \eq{eq:L2x2}, the $\omega$-dependence of the right--hand side has crucial consequences onto the dispersion of collective excitations. Of course, the usual Bogoliubov dispersion is recovered in the limit where high-$\omega$ perturbations are considered, so that $g_{\rm eff}$ recovers $g$. On the other hand, the static value \eq{eq:geff} for $g_{\rm eff}$ is recovered for stationary perturbations at $\omega=0$.


\subsection{Polaritons at rest $\mathbf{k}_p=0$}

Let us start from the $\mathbf{k}_p=0$ case.
In this case, the Bogoliubov matrix \eq{L3x3} is characterized by particle-hole  and parity symmetries, that combine in
\begin{equation}
  \mathscr{P} \mathscr{L}(\mathbf{ k }) = - \mathscr{L}(\mathbf{ k }) \mathscr{P}
\end{equation}
where
\begin{equation}
\mathscr{P} = \mathscr{K}
\begin{pmatrix}
\ 0 \  & 1 \  &  0 \ \\
\ 1 \  & 0  \ & 0  \ \\
\ 0 \  & 0 \ & 1 \
\end{pmatrix}
\end{equation}
and $\mathscr{K}$ stands for complex conjugation.
For a generic eigenvector $| \omega_{\mathbf{ k }} \rangle$ of $\mathscr{L}(\mathbf{ k })$ of eigenvalue $\omega$, this symmetry implies that
\begin{equation}
\mathscr{L}(\mathbf{ k })\, \mathscr{P} | \omega_{\mathbf{ k }}  \rangle = - \omega^* \mathscr{P} | \omega_{\mathbf{ k }} \rangle\,,
\end{equation}
i.e. that $\mathscr{P} | \omega_{\mathbf{ k }}  \rangle$
is itself an eigenvector of $\mathscr{L}(\mathbf{ k })$ of eigenvalue $-\omega^*$. This imposes the presence of pairs of eigenvectors with the same imaginary part and opposite real parts. Since the size of the matrix is three, this guarantees that at least one eigenvalue is purely imaginary. This mode can be interpreted as a reservoir branch $\omega^{R}(\mathbf{k})=-i \gamma^R(k) $, while the remaining two eigenvalues, corresponding to particle- and hole-like branches have general complex dispersions of the form $\omega^{\pm}(\mathbf{k}) = \pm \epsilon(k) - i {\gamma(k)}/{2}$.

Let us focus on the most relevant resonant case $\omega_p-\omega_0=\mu_T=gn_0+g_R n_R$ where the dispersion is expected to be gapless and sonic. In this regime, it is possible to obtain some analytical insight on the eigenvalue problem, which can be recast as
\begin{equation}
\left( \omega  + i\frac{\gamma}{2} \right)^2 = \frac{\hbar k^2}{2m}
 \left[ \frac{\hbar k^2}{2m} + 2 g_{\rm eff}(\omega) n_0    \right].
\end{equation}
At small $k$, this yields
\begin{equation}
 \omega^{\pm}(\mathbf{k}) = \pm c_s k - i  \gamma/2
 \label{eq:sonic}
\end{equation}
with
\begin{equation}
m c_s^2 = \hbar \mu_T + \frac{\gamma}{2 \gamma_R-\gamma} \hbar g_R n_R
\label{cx}
\end{equation}
In the fast reservoir limit $ \gamma_R  \gg \gamma$, the contribution of the reservoir is negligible and one recovers the usual speed of sound $m c_{s,T}^2 = \hbar \mu_T$ in terms of the total blue-shift $\mu_T$.

In the opposite limit $\gamma_R \ll  \gamma$, corresponding to the typical experimental conditions where the reservoir reacts on a much slower timescale~\cite{sarkar2010,walker2017,stepanov2019}, the speed of sound has the smaller value
\begin{equation}
m c_{s,0}^2 = \hbar \mu_T - \hbar g_R n_R = \hbar g n_0 \ .
\label{eq:true_cs}
\end{equation}
This means that, out of the total blue-shift $\mu_T$, only the component ($g n_0=g|\psi_0|^2$) due to the polaritons contributes to the speed of sound, while the one ($g_R n_R$) due to the incoherent reservoir only provides a global blue shift of the pumped mode.
This feature was experimentally observed in the pioneering experiment~\cite{stepanov2019} and is illustrated in the left panels of Fig.~\ref{fig:disp_kp}, showing the real and imaginary parts of the dispersion in panels (a) and (b), respectively. As expected, the cyan dashed lines in panel (a) indicate the sonic dispersion $\omega_s=\pm c_{s,0} k$ with the  speed of sound $c_{s,0}$  predicted by (\ref{eq:true_cs}) are in excellent agreement with the exact dispersion at low $k$'s. At higher $k$'s, the dispersion recovers the parabolic single-particle shape. As one can see in panel (b), the imaginary part of the reservoir mode (on the order of $\gamma_R$) remains always much smaller than the one of the sonic modes (on the order of $\gamma/2$).

While this picture is fully accurate when $\gamma$ is very much larger than $\gamma_R$, a subtle distinction must be done when $\gamma$ is larger but still somehow comparable to $\gamma_R$. In this regime, corrections in $\gamma_R/\gamma$ are important and one must distinguish the low-$k$ speed of sound set by \eq{cx} to the one at higher-$k$'s such that $|-i\omega(k) + \gamma/2| \ggg \gamma_R$, for which one exactly recovers \eq{eq:true_cs}. Also in this case, of course, the sonic behaviour is only visible up to the interaction energy $\hbar g n_0$, beyond which the dispersion recovers a single-particle behaviour~\footnote{This transition is determined by (the inverse of) the usual healing length, $\sqrt{\hbar^2/(g n_0 m)}$  computed including the polariton density only and the bare coupling $g$.}. The physical explanation is that at very small frequencies the reservoir can still (weakly) respond, while at higher $\omega$'s it behaves as a completely static background for the coherent field fluctuations. In order to clearly see the kink in the dispersion coming from distinction between $c_s$ and $c_{s,0}$, in Fig.~\ref{fig:true_cs} of the Appendix we tune $\gamma_R$ closer to $\gamma$.


In the intermediate case where $\gamma_R$ and $\gamma$ have comparable values and the blue-shift due to the reservoir is a significant fraction of $\mu_T$, the squared speed of sound $c_s^2$ predicted by \eq{cx} may becomes negative. This results in a flat Re$[\omega_{\pm}(\mathbf{k})]=0$ at small ${k}$ and a linear shape of the Im$[\omega_{\pm}({k})]$ starting from $-\gamma/2$. For larger ${k}$, the slope of the dispersion approaches the real-valued speed of sound $c_{s,0}$.
The usual sonic regime with a real-valued $c_s$ is found for higher values of the blueshift.

\subsection{Moving polaritons at finite $\mathbf{k}_p$}
\label{sec:Bogo_finite}

We conclude this section by extending the analysis to the case of a finite in-plane momentum $\mathbf{k_p}\neq 0$, which breaks parity. Therefore, the action of the $\mathscr{P}$ symmetry only entails
\begin{equation}
  \mathscr{P} \mathscr{L}(\mathbf{ k }) = - \mathscr{L}(-\mathbf{ k }) \mathscr{P}
\end{equation}
and relates eigenvectors at opposite $\mathbf{ k }$,
\begin{equation}
 \mathscr{L}(-\mathbf{ k })\, \mathscr{P} | \omega_{\mathbf{ k }}  \rangle = - \omega^* \mathscr{P} | \omega_{\mathbf{ k }} \rangle\,,
\end{equation}
that is $\mathscr{P} | \omega_{\mathbf{ k }}  \rangle$ is an eigenvector of $\mathscr{L}(-\mathbf{ k })$ of eigenvalue $-\omega^*$. This no longer implies the presence of a purely imaginary reservoir mode and the three branches are now strongly mixed as one can see in the right panels of Fig.~\ref{fig:disp_kp}. Note that the branches are colored here according to their nature at large wavevectors, while their mixing at small and intermediate $k$ complicates their classification. For instance, in the supersonic flow case considered here, the sonic mode with a wavevector $\mathbf{ k }$ directed in the upstream direction (that is, $k_x<0$) is strongly mixed with the reservoir. In panel (c), the Doppler-shifted sonic dispersions
\begin{equation}
 \omega=\pm c_{s,0} k +  \mathbf{v}_p \cdot \mathbf{ k } - i \gamma/2
 \label{eq:dopplersonic}
 \end{equation}
with the speed of sound (\ref{eq:true_cs}) and the flow speed $\mathbf{v}_p= \hbar \mathbf{k}_p/m$ (directed along the $x$ axis) are plotted as a dashed cyan line.
Note that this form of the Doppler shift is only accurate for small values of the momentum $k$, in contrast to the case with no reservoir where it holds for any $k$.

These concepts are further illustrated in the Appendix, where we plot three different response functions in the four cases with and without the incoherent reservoir and for a fluid at rest or in motion. In particular, one can see in the last plot that the brightness of one side of the upper branch is strongly reinforced in an experiment where Bogoliubov excitations are generated by a phononic white noise. This effect can only occur in the presence of the reservoir, otherwise the response functions would be simply rotated by the Doppler shift.
%

While the dispersions shown in Fig.~\ref{fig:disp_kp} are all dynamically stable, it is worth stressing that the presence of the reservoir can make a uniform flow at finite $k_p$ dynamically unstable, as signalled by a positive imaginary part of the dispersion.  With respect to panels (c-d) of Fig.~\ref{fig:disp_kp}, a slight increase of $\gamma_{inc}$ and thus of the reservoir fraction, or of the flow velocity $k_p$ will make the flow unstable by pushing the peaks in $\textrm{Im}[\omega]$ above zero. Similar modulational instabilities in the presence of a reservoir have been discussed in~\cite{wouters2007,bobrovska2014,bobrovska2017dynamical,baboux2018unstable}.


\section{Superfluidity  in the presence of an incoherent reservoir}
\label{sec:LandauReservoir}

\begin{figure*}[htbp]
\includegraphics[width=1.8\columnwidth]{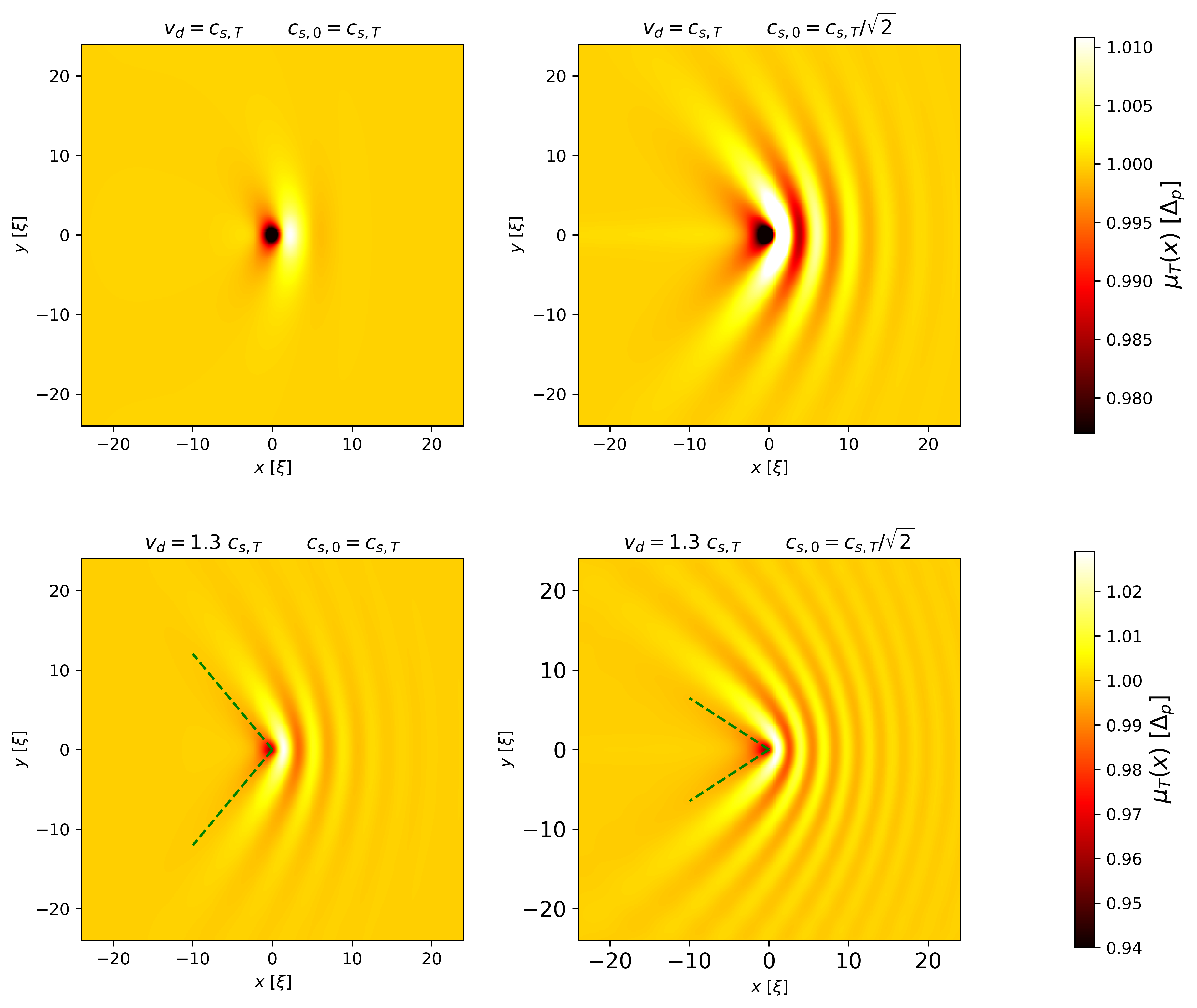}
\caption{Density modulation induced by a moving defect in the absence (left) and in the presence (right) of an incoherent reservoir. The total blue-shift $\mu_T$ is the same in all panels. The polariton fluid is at rest $k_p=0$ and the pump frequency is tuned at the resonant point $\Delta_p=\mu_T$. In the upper panels, the defect speed is chosen in the vicinity of the critical speed for superfluidity in the absence of incoherent reservoir, $v_d =  c_{s,T}$. In the lower panels, the defect speed is larger $v_d=1.3 \, c_{s,T}$. The dashed green lines in the upper panels indicate the Mach cone of angle $2\alpha$ expected from the chosen values of the flow $v_d$ and sound \eq{eq:true_cs} speeds, $\sin\alpha = c_{s,0}/v_d$.
Reservoir parameters are close to the ones estimated in  Ref.~\onlinecite{stepanov2019}, $g_R=2g$, $\gamma_R = 2\gamma_{inc} = 0.08 \,\gamma$. For thse values, the contributions of the polaritons and the incoherent reservoir to the blueshift are equal, $g_R n_R=g n_0=\mu_T/2$.
}
\label{fig:incoherent_density}
\end{figure*}

\begin{figure*}[htbp]
\includegraphics[width=2.0\columnwidth]{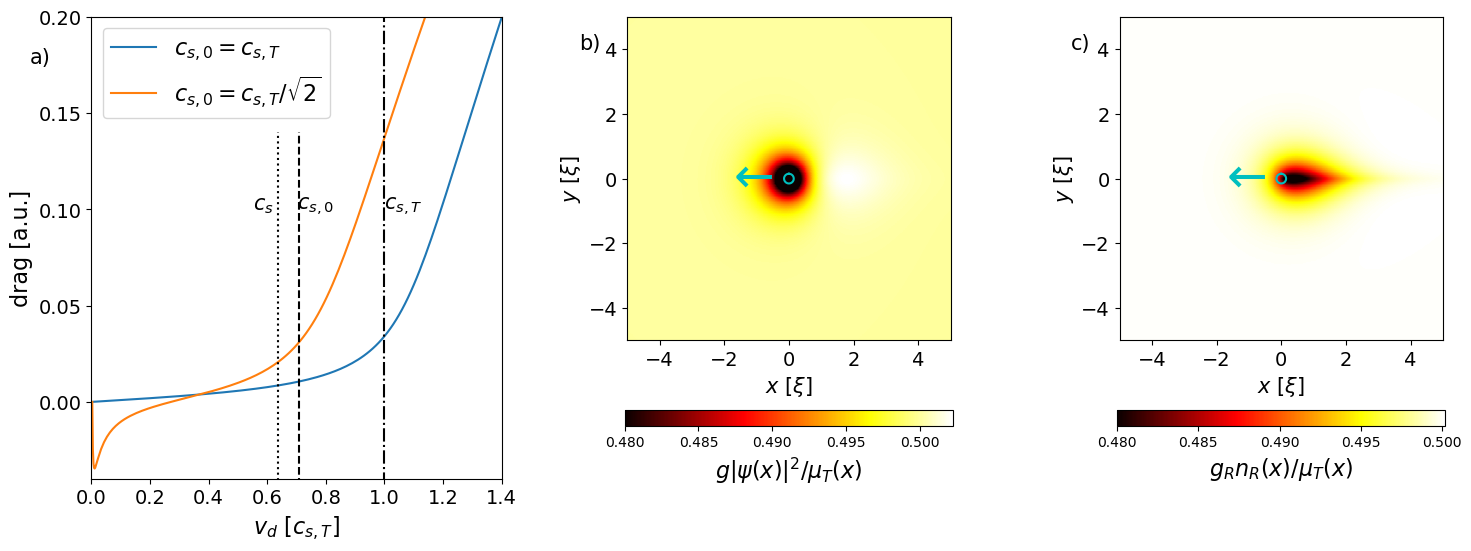}
\caption{
Critical speed for superfluidity in the presence of a reservoir, for a small and shallow defect moving in a polariton fluid pumped at the sonic point $\Delta_p=\mu_T$ and at rest $k_p=0$. (a) Drag force as a function of the defect velocity $v_d$ in the absence (blue) and in the presence (orange) of the incoherent reservoir. The force is here renormalized by the effective coupling $g_{\rm eff} F_d$, so to have a fair comparison of the two cases with and without reservoir.
The vertical lines confirm that in the former case the critical speed is at $c_{s,T}$, while in the latter case it is at $c_{s,0}$. An explanation for the negative drag at small $v_d$ is provided in panels (b,c) where $v_d = 0.02 c_{s,T}$ is taken. Panel (b) shows the polariton-induced component to the blueshift $g |\psi(x)|^2$. Panel (c) shows the incoherent reservoir contribution $g_R n_R(x)$. The defect consists of a gaussian perturbation indicated in the plot by the cyan circle of radius three times its width. The depletion of the (slow) reservoir density that it leaves behind it is partly  filled by the (faster) polariton. Same parameters as in the previous figures, namely $\gamma=0.2 \mu_T$, $g_R=2g$, $ \gamma_R = 2\gamma_{inc} = 0.08 \, \gamma$, so that $c_{s,0} =  c_{s,T}/\sqrt{2}$.
}
\label{fig:force}
\end{figure*}

In Sec.\ref{sec:GalileoBoost} we have seen that in the absence of reservoir, the generalized Gross-Pitaevskii equation \eq{eq:unboostedGPE} has specific invariance properties under Galilean boosts and in Sec.\ref{sec:LandauCriteria} we have shown that the superfluidity properties must be then the same in the two cases of a defect moving through a polariton fluid at rest and of a moving polariton fluid hitting a static defect: these two configurations represent in fact the same process seen in two different Galilean frames.
Correspondingly, since the (not invariant but) covariant coherent pump does not explicitely enter the linearized Bogoliubov calculation, the complex-valued  dispersion $\omega(\mathbf{k})$ simply gets Doppler-shifted $\omega(\mathbf{k})\rightarrow \omega(\mathbf{k})+\mathbf{k}\cdot \mathbf{v}$ when going to a reference frame moving at speed $\mathbf{v}$.

The situation is completely different in the presence of an incoherent reservoir, as described by the generalized dynamics of Eqs.(\ref{eq:psi}-\ref{eq:n_R}). This latter, in fact, defines a privileged frame of reference linked to the underlying semiconductor cavity structure. Such a feature is visible by comparing the Bogoliubov spectra shown in the left and right panels of Fig.\ref{fig:disp_kp}: even though the total blue-shift is the same in the two cases, the dispersions are markedly different in both the real and the imaginary parts.

In Fig.\ref{fig:incoherent_density}, we illustrate this breaking of Galilean invariance by looking at the effect of the incoherent reservoir on the density modulation pattern generated by a defect in motion through a fluid at rest.
As we expected and explicitly verified by numerical integration of the Gross-Pitaevskii equation (\ref{eq:psi}-\ref{eq:n_R}), a defect moving with constant velocity $\mathbf{v}_d$ in a fluid at rest with respect to the semiconductor substrate generates a pattern which is stationary in the frame of reference of the defect.

Therefore, within linear response to a shallow defect, it is possible to solve for the field perturbation in this frame by using the technique illustrated in Eq.~\eq{eq:linearized_GPE}. 
Since the reservoir equation in the defect frame \footnote{Remember that first-order equations for scalar quantities  require total time derivatives (also called convective derivatives) in order to be Galileo invariant, like the one for density in the incompressible Euler continuity equation.} reads
\begin{equation}
\partial_t n_R = -\mathbf{v}_d \cdot \nabla n_R - \gamma_R n_R + \gamma_{inc} |\psi|^2 \ ,
\end{equation}
the $\omega=0$ condition discussed in Sec.\ref{sec:LandauCriteria} allows for elimination of the reservoir via a momentum-dependent effective coupling
\begin{equation}
g_{\rm eff}(\mathbf{k}) =g + \frac{ \gamma_{inc}}{-i\mathbf{v}_d \cdot \mathbf{ k } + \gamma_R} g_R~.
\end{equation}
Notice that this procedure of imposing $\omega=0$ in the defect frame can be equivalently implemented in the lab frame by solving for $\omega=\mathbf{v}_d\cdot \mathbf{k}$; this is proven by expressing the defect potential  as $\delta V_{def}(\mathbf{x}-\mathbf{v}_dt) = \int d\mathbf{k}d\omega \ \delta(\omega-\mathbf{v}_d \cdot \mathbf{k}) \ \delta V_{def}(\mathbf{k}) e^{i\mathbf{k}\cdot\mathbf{x} -i\omega t}$, and similarly for the Ansatz of the field and reservoir.


For a fully coherent polariton fluid in the absence of a reservoir, the Galilean invariance holds and the physics only depends on the relative velocity of the fluid and the defect. As a result, the left panels of Fig.~\ref{fig:incoherent_density} equivalently represent the two cases of a fluid flowing against a static defect or of a moving defect in a fluid at rest.

On the basis of the discussion in the previous sections, it is natural to expect that the situation be completely different in the presence of an incoherent reservoir, which sets a privileged reference frame linked to the semiconductor matrix.
To start with, a pattern identical to the fully coherent case is  found for a static defect via the renormalized coupling (\ref{eq:geff}), as long as the  total blueshift is the same and no dynamical instabilities develop~\citep{amelio2019a}.
Instead, when it is the  defect to move in a polariton fluid at rest in the presence of a reservoir.  the density modulation pattern is shown in the right panels of Fig.~\ref{fig:incoherent_density}. These panels are plotted in the experimentally relevant $\gamma_R\ll \gamma$ regime for the same values of the speed $\mathbf{v}_d$ and the total interaction energy $\mu_T$ used in the left panels. It is apparent that the critical speed is strongly reduced, as expected from the Bogoliubov dispersion discussed in Sec.\ref{sec:Bogoliubov}. Moreover, the  shape of the density modulation profile shows a clear Mach cone  of angle $2\alpha$ with $\sin\alpha \simeq c_{s,0}/v_d$.

A more quantitative insight on the critical speed can be obtained looking at the plot of the friction force as a function of the defect speed for a polariton fluid at rest shown in Fig.\ref{fig:force}. The force is evaluated using \eq{eq:Fd} under the assumption that the defect only interacts with the coherent polaritons. Both in the absence (blue line) and in the presence (orange) of the reservoir, the friction force displays a clear threshold behaviour, losses being as usual~\cite{berceanu2012} responsible for a smoothening of the threshold. In contrast to the 1D case of Fig.\ref{fig:1Dsuper}, in the high-speed limit the force tends to the asymptotically linear dependence on $v_d$ predicted by~\cite{astrakharchik2004}.

As expected, the position of the threshold occurs at a markedly lower speed in the presence of the reservoir, at a value consistent with the effective speed of sound $c_{s,0}$. The fact that the critical speed is set by the effective high-$k$ speed of sound $c_{s,0}$ rather than by the low-$k$ value $c_s$ is physically understood by noting that the density modulation is peaked in $\mathbf{k}$-space at the intersection of the Bogoliubov dispersion with the $\omega=\mathbf{k}\cdot \mathbf{v}_d$ condition for the moving defect. A further confirmation of this statement can be found in Fig.~\ref{fig:true_cs}.a of the Appendix, where we show the same plot for a faster reservoir for which the distinction between $c_{s,0}$ and $c_s$ is more evident.

The origin of the peculiar negative value $F_d<0$ found in the presence of the reservoir is  illustrated in the panels (b,c) of Fig.\ref{fig:force}.
A very slow defect excites quasi--resonantly the reservoir branch of the dispersion, leaving in its wake a reservoir depletion, which is partially refilled by the faster polaritons. This results in an excess of polaritons behind the defect and, thus, to a negative drag.
Of course, the fact that the force tends to accelerate (rather than brake) the defect does not violate energy conservation, since we are dealing with a driven-dissipative system.

Coming back to the case of a defect at rest in a moving fluid, here the density modulation pattern is stationary in the frame of the semiconductor cavity structure, so the $\omega=0$ value of the effective interaction constant $g_{\rm eff}(\omega)$ is to be used. As we have discussed in the previous sections, this value recovers the interaction constant $g_{\rm eff}$ defined in \eq{eq:geff} that enters the expression for the total blue shift $\mu_T$, so that the critical speed for superfluidity is set by $c_{s,T}$ such that $mc_{s,T}^2=\hbar \mu_T$.
It is quite remarkable how this simple result holds independently of the relative magnitude of the polariton and reservoir contributions to this latter and of the details of the complex Bogoliubov dispersion in a moving fluid discussed in Sec.\ref{sec:Bogo_finite}. The only requirement is that the flow is dynamically stable for the chosen pump parameters.
This last subtle feature is the reason why the pioneering experiments in~\cite{amo2009} were in quantitative agreement with a theory that did not include the reservoir. For what concerns the dynamical experiments in~\cite{nardin2011}, instead, the quantitative agreement with the reservoir-less theory was guaranteed by the fact that the experiments were performed using a short pulse of coherent pump light, so that the reservoir density did not have time to build up.

\section{Conclusions}
\label{sec:Conclu}

In this work we have reported a detailed theoretical study of the effect of a reservoir of incoherent excitations on the superfluidity properties of polariton fluids in planar microcavities. In the absence of a reservoir, a  formal
Galilean tranformation relates the two situations of a fluid flowing against a static defect and of a defect moving in a fluid at rest. As a result, the dispersions of the Bogoliubov excitations are related by a simple Doppler shift and the density modulation pattern are identical in the two cases, as it normally happens in Galilean invariant fluids of material particles in free space.

On the contrary, the presence of the reservoir fixes a privileged laboratory reference frame linked to the semiconductor cavity structure. This breaking of Galilean invariance is visible in the Bogoliubov dispersion of the collective excitations in the fluid and in the density modulation pattern generated by a defect: while the effective speed of sound probed by a defect at rest is univocally determined by the total blue shift of the polariton modes as in the experiments of Ref.~\onlinecite{amo2009}, the one probed by a moving defect is significantly smaller and mostly determined by the polariton contribution to the blue shift. This results is of crucial importance to reconcile the historical demonstrations of polariton superfluidity in~\cite{amo2009,nardin2011} with the recent experiment in~\cite{stepanov2019}.

Beyond the microcavity polariton systems on which this article is focused, our results can be straightforwardly applied to other physical realizations of fluids of light such as photons propagating in cavityless nonlinear optical media~\cite{carusotto2014superfluid}. While a sort of Galilean invariance along the transverse plane holds for instantaneous Kerr-like nonlinearities~\cite{fontaine2018observation}, a strong breaking of Galilean invariance is in fact expected to occur when the optical nonlinearity has a thermal nature~\cite{vocke2015experimental}. This is a crucial feature that needs being duly taken into account when using quantum fluids of light as quantum simulators.

\section*{Acknowledgements}
 We are grateful to M. Wouters  for stimulating exchanges; I.A. thanks L. Giacomelli for continuous discussions on Bogoliubov formalism.
We acknowledge financial support from the European Union FET-Open grant ``MIR-BOSE'' (n. 737017), from the H2020-FETFLAG-2018-2020 project ``PhoQuS'' (n.820392), and from the Provincia Autonoma di Trento. All numerical calculations were performed using the Julia Programming Language \cite{julia}.

\bibliography{bibliography}

\vspace*{0.1in}

\section*{Appendix}
\subsection*{ Dynamically relevant speed of sound}

\begin{figure}[htbp]
\includegraphics[width=0.64\columnwidth]{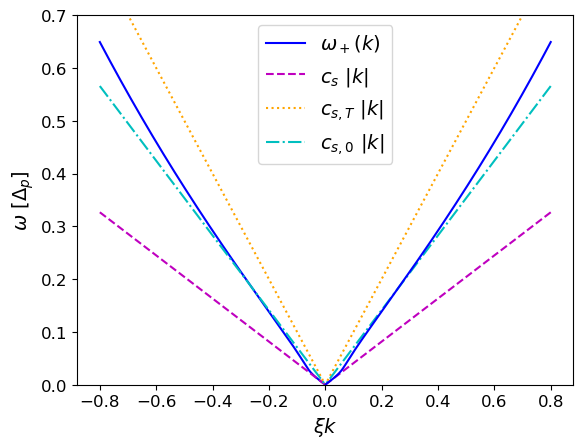}
\includegraphics[width=0.64\columnwidth]{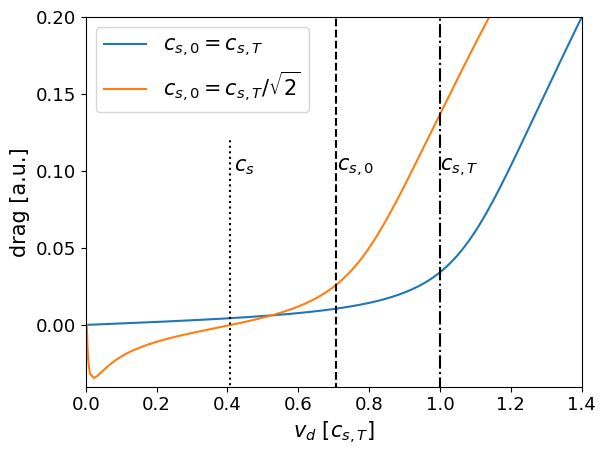}
\caption{(upper) 
Dispersion of the collective excitations of  a polariton fluid at the acoustic point and in the presence of reservoir 
with parameters tuned so to make clear the distinction between the three definitions of the speed of sound: $c_s$ is the slope of the dispersion at $k=0$, $c_{s,T}$ takes into account the total blueshift and it is correct in the adiabatic limit, $c_{s,0}$ is only due to the self-interaction of the coherent polariton fluid. This last one  turns out to be the critical velocity when superfluidity is considered, e.g.  by computing the drag force, as depicted in the lower panel. Here the blue line is in the absence of the reservoir and orange  in the presence of a reservoir such that  $\mu_T = 2 g|\psi_0|^2$. 
}\label{fig:true_cs}
\end{figure}

In the main text  we take for the reservoir relaxation rate  the value $\gamma_R = 0.08 \, \gamma$ directly estimated in \cite{stepanov2019}, and choose $g_R=2g, \gamma_{inc}=\gamma_R/2$, so to have half of the total blue-shift due to the reservoir and half to the coherent part of the fluid. With these parameters $c_{s,0}$ is quite close to $c_{s}$, which determines the slope of the dispersion only at very low momenta. Indeed, in Fig.~\ref{fig:disp_kp}.a we plot only  the comparison with $c_{s,0}$. Also the dynamical response to a moving defect is mainly determined by $c_{s,0}$, see Fig.~\ref{fig:force}.a. 
In order to better highlight the crucial distinction between $c_{s,0}$ and $c_{s}$, in this paragraph we set
$\gamma_R = 2\gamma_{inc} = 0.2 \, \gamma$. 
Doing so, it is well visible in Fig.~\ref{fig:true_cs}.a that $c_s$ defines the slope of the dispersion in the immediate proximity of  $k=0$, but very soon $c_{s,0}$ gets more relevant; when the wavevector exceeds the inverse of the healing length the parabolic single particle character of the dispersion dominates.
Also for the superfluidity properties the dynamically relevant critical velocity is clearly $c_{s,0}$, as probed by the drag force in Fig.~\ref{fig:true_cs}.b.

\begin{figure*}[]
\includegraphics[width=1.85\columnwidth]{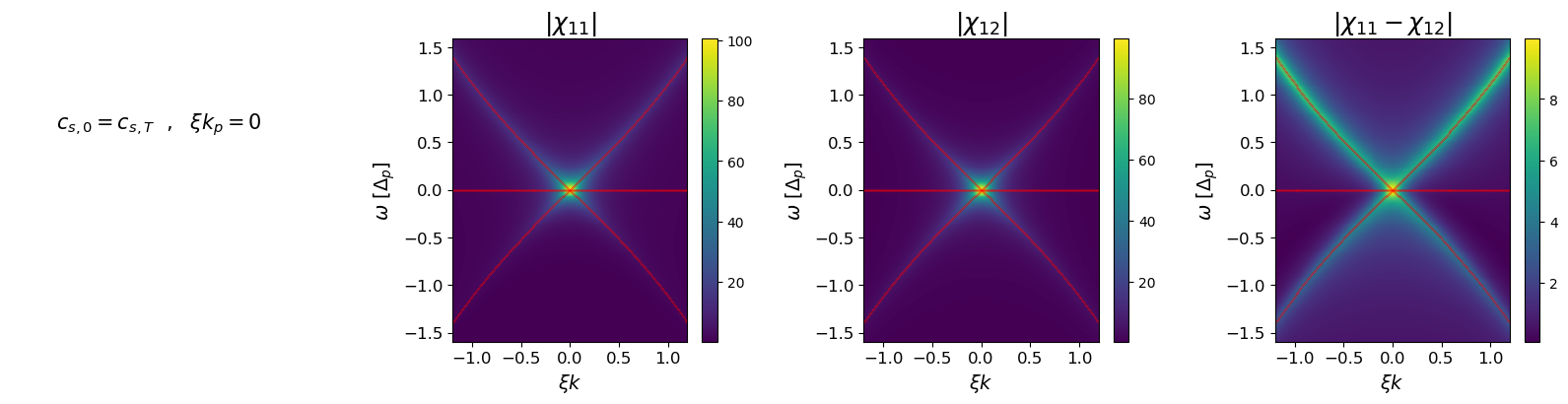}
\includegraphics[width=1.85\columnwidth]{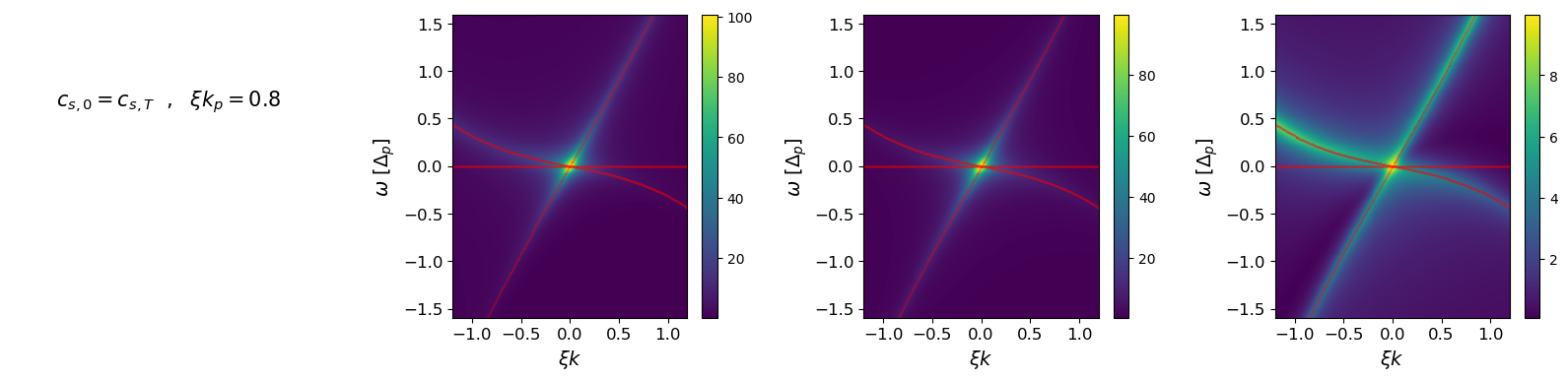}
\includegraphics[width=1.85\columnwidth]{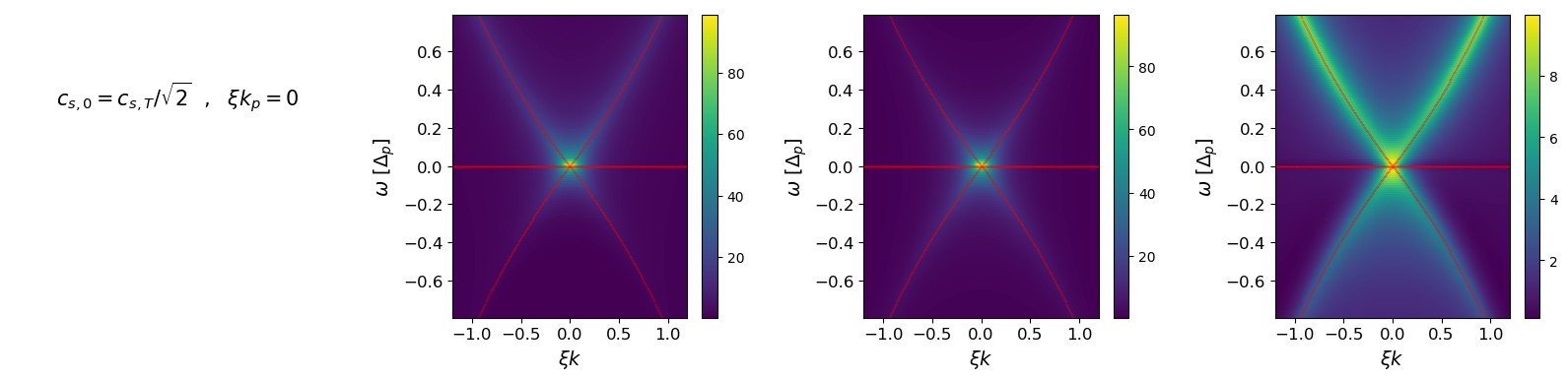}
\includegraphics[width=1.85\columnwidth]{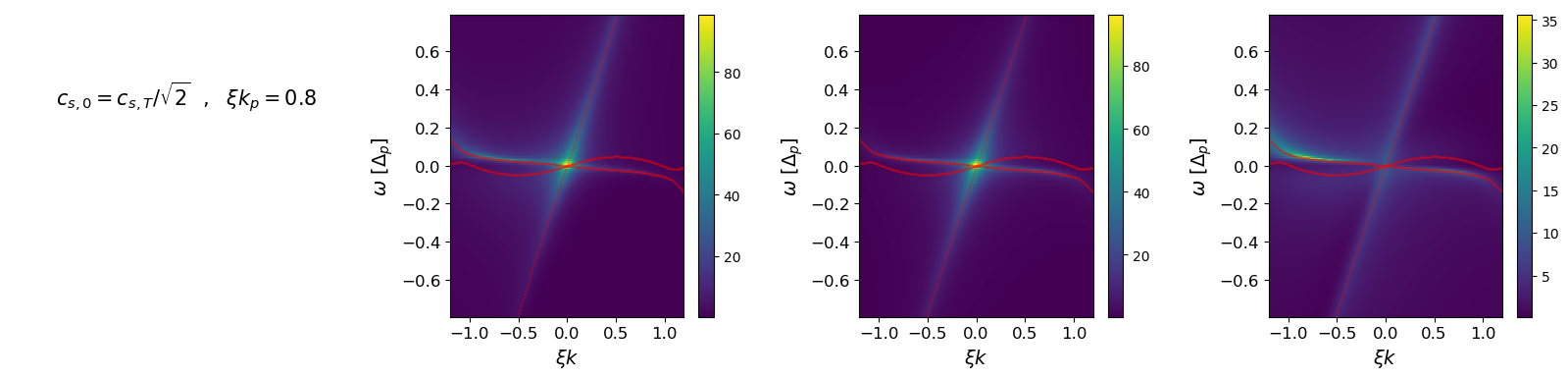}
\caption{From left to right: color plots of  $|\chi_{11}|, |\chi_{12}|, |\chi_{11} - \chi_{12}|$ as functions of $(k,\omega)$. From top to bottom: without reservoir and  $k_p=0$, without reservoir and  $\xi k_p=0.8$, with reservoir and $k_p=0$, with reservoir and $\xi k_p=0.8$. The reservoir parameters are the same as in Fig.~\ref{fig:disp_kp}. In particular, looking at the last column, it is clear that having both $\mathbf{k}_p \neq 0$ and a reservoir allows for having different luminescence (as generated by phononic white noise) on the left and right particle branches, while  the colorplot is only rigidly rotated according to the Doppler shift if the reservoir is absent.
}
\label{fig:chis}
\end{figure*}

\subsection*{Response functions}

The dynamic response function is defined as 
\begin{equation}
\chi(\mathbf{k},\omega) = \frac{1}{\omega - \mathscr{L}(\mathbf{k})} .
\end{equation}
Physically, $\chi_{11}(\mathbf{k},\omega)$ is the response to a probe at $(\mathbf{k}_p + \mathbf{k}, \omega_p + \omega)$ measured at the probe momentum and frequency, while $\chi_{12}$ describes the response in a four-wave mixing setup; finally,  $\chi_{11}-\chi_{12}$ was considered in \cite{stepanov2019} and represents the susceptibility to scattering with  phonons (or to any real field that couples to the polariton density).
In Fig.~\ref{fig:chis} we plot these quantities (from left to right), for four different situations. When a  polariton fluid is considered in the absence of  reservoir, Galilean invariance ensures that the physical susceptibility of a fluid at rest (first row) gets rigidly Doppler rotated when setting the fluid into motion  (second row).
In the very last plot, which refers to the case with a resevoir, the left particle branch is instead much brighter than the right one, which is only possible because Galilean invariance is broken.

\end{document}